\documentclass[12pt]{article}
\usepackage{amsmath}
\usepackage{color}
\usepackage{amssymb}
\usepackage{bm}

\makeatletter\@addtoreset{equation}{section}
\makeatother
\allowdisplaybreaks[1]
\begin{document}
\begin{titlepage}

\begin{flushright}
\phantom{preprint no.}
\end{flushright}
\vspace{0.5cm}
\begin{center}
{\Large \bf
Gravitational-wave equation in\\
\vspace{2mm}
effective one-body background for spinless binary
}
\lineskip .75em
\vskip0.5cm
{\large Ya Guo${}^{1}$, Hiroaki Nakajima${}^{2}$ and Wenbin Lin${}^{1,\,2,\,*}$}
\vskip 2.5em
${}^{1}$ {\normalsize\it School of Physical Science and Technology, Southwest Jiaotong University, \\ Chengdu, 610031, China\\
}
\vskip 1.0em
${}^{2}$ {\normalsize\it School of Mathematics and Physics, University of South China, \\ Hengyang, 421001, China\\}
\vskip 1.0em
${}^{*}$ {\normalsize\it Email: lwb@usc.edu.cn\\}
\vskip 1.0em
\vskip 3.0em
\end{center}
\begin{abstract}
We construct the gravitational-wave equation in the background of
the effective one-body system for the spinless binary, which is in general available with
the spherically symmetric background as well. The gauge conditions are given in terms of the
metric perturbation.
\end{abstract}
\end{titlepage}

\section{Introduction}
The direct observation of gravitational waves by LIGO and Virgo~\cite{Abbott:2016blz} has opened the new era of cosmology.
The binary system such as two black holes is a good object to observe the gravitational waves,
where the analytical calculation is also possible.
One of the calculation method of the gravitational waves radiated from the binary system
is the theory of post-Newtonian approximations \cite{PoissonWill}.
On the other hand, one can also use the black hole perturbation theory \cite{Mino:1997bx},
which is useful for the case of the extreme mass-ratio inspiral (EMRI).
The advantage of this method is that there is no divergence and one can calculate the observable quantity
at very high post-Newtonian order \cite{Fujita:2014eta}.

The reason why the black-hole perturbation theory is available for EMRI is because the field of one object is regarded
as the background and the other body is regarded as the test particle.
If one can map the two-body dynamics into the dynamics of the test particle
in some appropriate background \textit{exactly}, then the calculation beyond the EMRI approximation would be possible
in terms of the black-hole perturbation theory. This approach is called the effective one-body (EOB) dynamics
\cite{Buonanno:2000ef,Damour:2016gwp}.
In Newtonian limit, it is well-known that the effect of the two-body dynamics can exactly be included by just replacing
the light (heavy) mass in the EMRI approximation with the reduced (total) mass. However when the correction of
general relativity is included, the correspondence between the two-body dynamics and the dynamics of the test particle
in some background becomes complicated. It turns out that the Hamiltonians of the two dynamics are related
by a very nontrivial way \cite{Buonanno:2000ef,Damour:2016gwp}.

The gravitational-wave equation in the EOB background has been studied in \cite{Jing:2021ahx}
using the Newman-Penrose formalism \cite{Newman:1961qr}, which is a natural extension of the method to
obtain the Teukolsky equation \cite{Teukolsky:1973ha} in the Schwarzschild spacetime.
It is obtained for some special cases of the background and is restricted to the even-parity mode.
Later the same group obtained the equation both for the even- and odd-parity modes using the different choices
of the gauge conditions \cite{Jing:2022vks}. %
The wave equation from the metric perturbation in the generally spherically symmetric background has also been
studied in \cite{Lenzi:2021wpc,Liu:2022csl}.

Inspired by the work of Jing et al.~\cite{Jing:2021ahx}, in this paper, we show that the gravitational-wave equation for both the even- and odd-parity modes can be obtained using the gauge conditions taken in their work. Moreover our formalism can be applicable for more general spherically symmetric backgrounds.

The reminder of this work is organized as follows: in section 2, we briefly review the EOB dynamics.
In section 3, we consider the wave equation for the perturbed Weyl scalars.
In section 4, we discuss the gauge condition. In section 5, we derive the explicit gravitational-wave equation.
Section 6 is devoted to summary and discussion.

\section{EOB dynamics}

The EOB system for the spinless binary is first introduced in \cite{Buonanno:2000ef} in the post-Newtonian formalism.
The effective background metric $g_{\mu\nu}^{\mathrm{eff}}$ is taken as the spherically symmetric form:
\begin{gather}
ds^{2}_{\mathrm{eff}}=g_{\mu\nu}^{\mathrm{eff}}dx^{\mu}dx^{\nu}
=\mathcal{A}(r)dt^{2}-\mathcal{B}(r)dr^{2}-\mathcal{C}(r)r^{2}(d\theta^{2}+\sin^{2}\theta d\varphi^{2}), \label{bg}
\end{gather}
where the function $\mathcal{C}(r)$ can be freely chosen by the coordinate transformation for the radial coordinate $r$,
and we here choose the Schwarzschild coordinate corresponding to $\mathcal{C}(r)\equiv 1$.
The explicit forms of $\mathcal{A}(r)$ and $\mathcal{B}(r)$ can be determined by the comparison with the two-body dynamics.
For the Newman-Penrose formalism \cite{Newman:1961qr}, we will take the corresponding null tetrad basis
\begin{align}
l_{\mu}^{A}dx^{\mu}&=dt-\frac{\mathcal{D}(r)}{\mathcal{A}(r)}dr, &
n_{\mu}^{A}dx^{\mu}&=\frac{\mathcal{A}(r)}{2}dt+\frac{\mathcal{D}(r)}{2}dr, \notag\\
m_{\mu}^{A}dx^{\mu}&=-\frac{r}{\sqrt{2}}(d\theta+i\sin\theta d\varphi), &
\bar{m}_{\mu}^{A}dx^{\mu}&=-\frac{r}{\sqrt{2}}(d\theta-i\sin\theta d\varphi),
\end{align}
where $\mathcal{D}(r)=\sqrt{\mathcal{A}(r)\mathcal{B}(r)}$. The suffix $A$ denotes the background quantities.
The null tetrads satisfy the orthonormal condition as
\begin{align}
l_{\mu}^{A}n^{\mu}_{A}=1,\quad m_{\mu}^{A}\bar{m}^{\mu}_{A}=-1,
\end{align}
and the other inner products vanish.
From the tetrad basis, one can compute the spin coefficients, the components of the Ricci tensor and the Weyl scalars as
\begin{gather}
\kappa^{A}=\nu^{A}=\sigma^{A}=\lambda^{A}=\pi^{A}=\tau^{A}=\epsilon^{A}=0, \label{bg11}\\
\rho^{A}=-\frac{1}{r\mathcal{D}}, \quad \mu^{A}=-\frac{\mathcal{A}}{2r\mathcal{D}},
\quad \gamma^{A}=\frac{\mathcal{A}'}{4\mathcal{D}}, \quad \alpha^{A}=-\beta^{A}=-\frac{\cot\theta}{2\sqrt{2}r}, \\
\Phi_{01}^{A}=\Phi_{10}^{A}=\Phi_{02}^{A}=\Phi_{20}^{A}=\Phi_{12}^{A}=\Phi_{21}^{A}=0, \\
\Phi_{00}^{A}=-\frac{\mathcal{D}'}{r\mathcal{D}^{3}}, \quad \Phi_{22}^{A}=-\frac{\mathcal{A}^{2}\mathcal{D}'}{4rD^{3}}, \\
\Phi_{11}^{A}=-\frac{1}{8r^{2}\mathcal{D}^{3}}\Bigl[2\mathcal{D}^{3}-2\mathcal{A}\mathcal{D}
-r^{2}(\mathcal{A}'\mathcal{D}'-\mathcal{A}''\mathcal{D})\Bigr], \\
\Lambda^{A}=\frac{1}{24r^{2}\mathcal{D}^{3}}\Bigl[-2\mathcal{D}^{3}-r^{2}\mathcal{A}'\mathcal{D}'
+2\mathcal{A}(\mathcal{D}-2r\mathcal{D}')+r\mathcal{D}(4\mathcal{A}'+r\mathcal{A}'')\Bigr], \\
\Psi_{0}^{A}=\Psi_{1}^{A}=\Psi_{3}^{A}=\Psi_{4}^{A}=0, \label{typeD}\\
\Psi_{2}^{A}=\frac{1}{12r^{2}\mathcal{D}^{3}}\Bigl[2(\mathcal{A}\mathcal{D}+r\mathcal{A}\mathcal{D}'-\mathcal{D}^{3})-r\mathcal{A}'(2\mathcal{D}+r\mathcal{D}')+r^{2}\mathcal{A}''\mathcal{D}\Bigr], \label{bg18}
\end{gather}
where the prime denotes the ordinary derivative with respect to $r$.
One can find that the background belongs the petrov type D from \eqref{typeD},
but is not in the vacuum since there are nonvanishing components of the Ricci tensor.
This type D property is important to derive the gravitational-wave equation in the next section.
Note that for the special case $\mathcal{D}(r)\!=\! 1$, which was taken in \cite{Jing:2021ahx,Jing:2022vks}, we have
\begin{gather}
\Phi_{00}^{A}=\Phi_{22}^{A}=0, \label{0022}
\end{gather}
and the nonvanishing quantities in the above becomes simplified as
\begin{gather}
\rho^{A}=-\frac{1}{r}, \quad \mu^{A}=-\frac{\mathcal{A}}{2r}, \quad \gamma^{A}=\frac{\mathcal{A}'}{4},
\quad \alpha^{A}=-\beta^{A}=-\frac{\cot\theta}{2\sqrt{2}r}, \label{bg21}\\
\Phi_{11}^{A}=-\frac{1}{8r^{2}}(2-2\mathcal{A}+r^{2}\mathcal{A}''), \\
\Lambda^{A}=\frac{1}{24r^{2}}(-2+2\mathcal{A}+4r\mathcal{A}'+r^{2}\mathcal{A}''), \\
\Psi_{2}^{A}=\frac{1}{12r^{2}}(-2+2\mathcal{A}-2r\mathcal{A}'+r^{2}\mathcal{A}''), \label{bg24}
\end{gather}
Note that when we choose $\mathcal{A}=1-2M/r$ and $\mathcal{D}=1$, the background \eqref{bg} is reduced to
the Schwarzschild spacetime.

The EOB Hamiltonian $H_{\mathrm{eff}}$ can be determined from the geodesic motion under the background \eqref{bg}.
The action $S$ satisfies the Hamilton-Jacobi equation
\begin{gather}
g^{\mu\nu}_{\mathrm{eff}}P_{\mu}P_{\nu}-m_{0}^{2}=0, \label{HJ}
\end{gather}
where $P_{\mu}=\partial S/\partial x^{\mu}$ is the momentum.
$m_{0}$ is the mass of the test particle and is matched as the reduced mass in the two-body dynamics.
From \eqref{HJ}, $H_{\mathrm{eff}}$ is computed as
\begin{gather}
H_{\mathrm{eff}}=m_{0}\sqrt{\mathcal{A}
\left(1+\frac{\mathcal{A}P_{r}^{2}}{m_{0}^{2}\mathcal{D}^{2}}+\frac{P_{\varphi}^{2}}{m_{0}^{2}r^{2}}\right)}\ ,
\end{gather}
where the motion plane is fixed on $\theta=\pi/2$ due to the spherical symmetry.
The effective Hamiltonian $H_{\mathrm{eff}}$ and the  real two-body Hamiltonian $H_{\mathrm{real}}$
are compared by matching the masses and the action variables.
It turns out that the two Hamiltonians are related by a rather nontrivial way as \cite{Buonanno:2000ef,Damour:2016gwp}
\begin{gather}
H_{\mathrm{real}}=M_{0}\sqrt{1+\frac{2m_{0}}{M_{0}}\left(\frac{H_{\mathrm{eff}}}{m_{0}}-1\right)}\ ,
\end{gather}
where $M_{0}$ is
the total mass in the two-body dynamics. The functions $\mathcal{A}(r)$ and $\mathcal{D}(r)$ are obtained as
\cite{Buonanno:2000ef}
\begin{gather}
\mathcal{A}(r)=1-\frac{2M_{0}}{r}+\frac{2m_{0}}{M_{0}}\left(\frac{M_{0}}{r}\right)^{3}+\cdots, \quad
\mathcal{D}(r)=1-\frac{3m_{0}}{M_{0}}\left(\frac{M_{0}}{r}\right)^{2}+\cdots.
\end{gather}
However hereafter we leave $\mathcal{A}(r)$ and $\mathcal{D}(r)$ arbitrarily. Because of that, the metric \eqref{bg}
takes the most general form of the spherically symmetric background, which can also be applied to other kinds of the background.
Moreover we will see later that when $\mathcal{D}(r)\!=\! 1$, the gravitational-wave equation and the gauge condition
becomes simplified drastically.

\section{Wave equation for perturbed Weyl scalars}

It has been shown that the background \eqref{bg} is classified as
the nonvacuum Petrov type D background, which is useful to derive the
gravitational-wave equation for the perturbed Weyl scalars using the Newman-Penrose formalism,
as in the Teukolsky equation \cite{Teukolsky:1973ha} in the Schwarzschild and the Kerr background.

We begin with the following equations in Newman-Penrose formalism:
\begin{align}
&(\delta+4\beta-\tau)\Psi_{4}-(\Delta+4\mu+2\gamma)\Psi_{3}+3\nu\Psi_{2} \notag\\
&\qquad =(\bar{\delta}-\bar{\tau}+2\bar{\beta}+2\alpha)\Phi_{22}-(\Delta+2\gamma+2\bar{\mu})\Phi_{21}
-2\lambda\Phi_{12}+2\nu\Phi_{11}+\bar{\nu}\Phi_{20}, \label{eq1}\\
&(D+4\epsilon-\rho)\Psi_{4}-(\bar{\delta}+4\pi+2\alpha)\Psi_{3}+3\lambda\Psi_{2} \notag\\
&\qquad =(\bar{\delta}-2\bar{\tau}+2\alpha)\Phi_{21}-(\Delta+2\gamma-2\bar{\gamma}+\bar{\mu})\Phi_{20}
+\bar{\sigma}\Phi_{22}-2\lambda\Phi_{11}+2\nu\Phi_{10}, \label{eq2}\\
&(\Delta+\mu+\bar{\mu}+3\gamma-\bar{\gamma})\lambda-(\bar{\delta}+\pi-\bar{\tau}+\bar{\beta}+3\alpha)\nu+\Psi_{4}=0.
\label{eq3}
\end{align}
We split all the quantities in the above into the background part $(A)$ and the perturbation part $(B)$,
\textit{e.\,g.} $\Psi_{4}=\Psi_{4}^{A}+\Psi_{4}^{B}$, etc.
Now we have to take into account the case where $\Phi_{22}^{A}$ is nonvanishing, then the background part
of the equation becomes nontrivial, \textit{i.\,e.}
\begin{gather}
(\bar{\delta}-\bar{\tau}+2\bar{\beta}+2\alpha)^{A}\Phi_{22}^{A}=0,
\end{gather}
has to be satisfied\footnote{Here the superscript $A$ $(B)$ on the parentheses denotes that all the quantities
and the operators inside the parentheses are in the background (perturbation).},
and one can confirm that it is indeed the case. The part of the first-order perturbation in \eqref{eq1}--\eqref{eq3}
becomes\footnote{Here we will not use $\pi^{A}=\tau^{A}=\epsilon^{A}=0$ from the beginning and keep them for a while,
which would be useful to extend the result into the spinning case.}
\begin{align}
&(\delta+4\beta-\tau)^{A}\Psi_{4}^{B}-(\Delta+4\mu+2\gamma)^{A}\Psi_{3}^{B}+3\nu^{B}\Psi_{2}^{A} \notag\\
&\qquad =(\bar{\delta}-\bar{\tau}+2\bar{\beta}+2\alpha)^{B}\Phi_{22}^{A}
+(\bar{\delta}-\bar{\tau}+2\bar{\beta}+2\alpha)^{A}\Phi_{22}^{B} \notag\\
&\qquad\quad\  -(\Delta+2\gamma+2\bar{\mu})^{A}\Phi_{21}^{B}+2\nu^{B}\Phi_{11}^{A}, \label{eq01}\\
&(D+4\epsilon-\rho)^{A}\Psi_{4}^{B}-(\bar{\delta}+4\pi+2\alpha)^{A}\Psi_{3}^{B}+3\lambda^{B}\Psi_{2}^{A} \notag\\
&\qquad =(\bar{\delta}-2\bar{\tau}+2\alpha)^{A}\Phi_{21}^{B}-(\Delta+2\gamma-2\bar{\gamma}+\bar{\mu})^{A}\Phi_{20}^{B}
+\bar{\sigma}^{B}\Phi_{22}^{A}-2\lambda^{B}\Phi_{11}^{A}, \label{eq02}\\
&(\Delta+\mu+\bar{\mu}+3\gamma-\bar{\gamma})^{A}\lambda^{B}-(\bar{\delta}+\pi-\bar{\tau}+\bar{\beta}+3\alpha)^{A}\nu^{B}
+\Psi_{4}^{B}=0. \label{eq03}
\end{align}
Again, when  $\Phi_{22}^{A}$ is nonvanishing, the first term in the right hand side in \eqref{eq01} appears,
and more perturbed quantities $\bar{\tau}^{B}$, $\bar{\beta}^{B}$ and $\alpha^{B}$ contribute to the equation,
compared with the vacuum case.
In order to reduce the number of those quantities, we require that this term should vanish by the
gauge condition;
\begin{gather}
\Xi_{22}\equiv(\bar{\delta}-\bar{\tau}+2\bar{\beta}+2\alpha)^{B}\Phi_{22}^{A}=0, \label{gauge1}
\end{gather}

Now we will obtain the wave equation for $\Psi_{4}^{B}$ in a similar way with the method
used to derive the Teukolsky equation. First we show the following commutation relation of the
differential operators:
\begin{align}
&\left[\Delta+(p+1)\gamma-\bar{\gamma}-q\mu+\bar{\mu}\right]^{A}(\bar{\delta}+p\alpha-q\pi)^{A} \notag\\
&\qquad -\left[\bar{\delta}+(p+1)\alpha+\bar{\beta}-\bar{\tau}-q\pi\right]^{A}(\Delta+p\gamma-q\mu)^{A} \notag\\
&=\nu^{A}D^{A}-\lambda^{A}\delta^{A}-p\left[(\beta+\tau)\lambda-(\rho+\epsilon)\nu+\Psi_{3}\right]^{A} \notag\\
&\qquad +q\left[-D\nu+\delta\lambda+(\bar{\pi}+\tau+3\beta-\bar{\alpha})\lambda
-(3\epsilon+\bar{\epsilon}+\rho-\bar{\rho})\nu+2\Psi_{3}\right]^{A} \notag\\
&=0, \label{com}
\end{align}
where $p$ and $q$ are arbitrary constants and we have used $\nu^{A}=\lambda^{A}=\Psi_{3}^{A}=0$.
We operate $(\Delta+3\gamma-\bar{\gamma}+4\mu+\bar{\mu})^{A}$ to \eqref{eq02}
and $(\bar{\delta}+3\alpha+\bar{\beta}-\bar{\tau}+4\pi)^{A}$ to \eqref{eq01}, and then subtract one equation
from the other. The terms with $\Psi_{3}^{B}$ cancel by \eqref{com} with $p=2$, $q=-4$
and the remaining becomes
\begin{align}
&\left[(\Delta+3\gamma-\bar{\gamma}+4\mu+\bar{\mu})(D+4\epsilon-\rho)
-(\bar{\delta}+3\alpha+\bar{\beta}-\bar{\tau}+4\pi)(\delta+4\beta-\tau)\right]^{A}\Psi_{4}^{B} \notag\\
&\qquad {}+3\Psi_{2}^{A}\left[(\Delta+3\gamma-\bar{\gamma}+4\mu+\bar{\mu})^{A}\lambda^{B}
-(\bar{\delta}+3\alpha+\bar{\beta}-\bar{\tau}+4\pi)^{A}\nu^{B}\right] \notag\\
&\qquad {}+3\lambda^{B}\Delta^{A}\Psi_{2}^{A}-3\nu^{B}\bar{\delta}^{A}\Psi_{2}^{A} \notag\\
&=T_{4}+(\Delta+3\gamma-\bar{\gamma}+4\mu+\bar{\mu})^{A}(\bar{\sigma}^{B}\Phi_{22}^{A})
-2\lambda^{B}\Delta^{A}\Phi_{11}^{A}-2\nu^{B}\bar{\delta}^{A}\Phi_{11}^{A} \notag\\
&\qquad {}-2\Phi_{11}^{A}\left[(\Delta+3\gamma-\bar{\gamma}+4\mu+\bar{\mu})^{A}\lambda^{B}
+(\bar{\delta}+3\alpha+\bar{\beta}-\bar{\tau}+4\pi)^{A}\nu^{B}\right], \label{eq04}
\end{align}
where $T_{4}$ is defined by
\begin{align}
T_{4}&=(\Delta+3\gamma-\bar{\gamma}+4\mu+\bar{\mu})^{A}
\left[(\bar{\delta}-2\bar{\tau}+2\alpha)^{A}\Phi_{21}^{B}
-(\Delta+2\gamma-2\bar{\gamma}+\bar{\mu})^{A}\Phi_{20}^{B}\right]
\notag\\
&\quad {}-(\bar{\delta}+3\alpha+\bar{\beta}-\bar{\tau}+4\pi)^{A}
\left[(\bar{\delta}-\bar{\tau}+2\bar{\beta}+2\alpha)^{A}\Phi_{22}^{B}
-(\Delta+2\gamma+2\bar{\mu})^{A}\Phi_{21}^{B}\right].
\end{align}
For the third line in \eqref{eq04}, we have
\begin{align}
\Delta^{A}\Psi_{2}^{A}&=-3\mu^{A}\Psi_{2}^{A}-2\mu\Phi_{11}^{A}
-(D-\bar{\rho}+2\epsilon+2\bar{\epsilon})^{A}\Phi_{22}^{A}-2\Delta^{A}\Lambda^{A}, \\
\bar{\delta}^{A}\Psi_{2}^{A}&=-3\pi^{A}\Psi_{2}^{A}+2\pi^{A}\Phi_{11}^{A}-2\bar{\delta}^{A}\Lambda^{A},
\end{align}
and then substituting the above into \eqref{eq04}, we get
\begin{align}
&\left[(\Delta\!+\!3\gamma\!-\!\bar{\gamma}\!+\!4\mu+\bar{\mu})(D\!+\!4\epsilon\!-\!\rho)
\!-\!(\bar{\delta}+3\alpha+\bar{\beta}-\bar{\tau}+4\pi)(\delta+4\beta-\tau)\right]^{A}\Psi_{4}^{B} \notag\\
&\qquad {}+3\Psi_{2}^{A}\left[(\Delta+3\gamma-\bar{\gamma}+\mu+\bar{\mu})^{A}\lambda^{B}
-(\bar{\delta}+3\alpha+\bar{\beta}-\bar{\tau}+\pi)^{A}\nu^{B}\right] \notag\\
&=T_{4}+(\Delta+3\gamma-\bar{\gamma}+4\mu+\bar{\mu})^{A}(\bar{\sigma}^{B}\Phi_{22}^{A})
-2\lambda^{B}\Delta^{A}\Phi_{11}^{A}-2\nu^{B}\bar{\delta}^{A}\Phi_{11}^{A} \notag\\
&\qquad {}-2\Phi_{11}^{A}\left[(\Delta+3\gamma-\bar{\gamma}+\mu+\bar{\mu})^{A}\lambda^{B}
+(\bar{\delta}+3\alpha+\bar{\beta}-\bar{\tau}+\pi)^{A}\nu^{B}\right] \notag\\
&\qquad {}+3\lambda^{B}(D-\bar{\rho}+2\epsilon+2\bar{\epsilon})^{A}\Phi_{22}^{A}
+6\lambda^{B}\Delta^{A}\Lambda^{A}-6\nu^{B}\bar{\delta}^{A}\Lambda^{A}.
\label{eq05}
\end{align}
The second line in \eqref{eq05} becomes $-3\Psi_{2}^{A}\Psi_{4}^{B}$ using \eqref{eq03},
and there is also similar terms in the fourth line but the relative sign is positive.
We now require more gauge conditions as
\begin{gather}
\lambda^{B}=\sigma^{B}=0. \label{gauge2}
\end{gather}
Then the fourth line in \eqref{eq05} becomes $-2\Phi_{11}^{A}\Psi_{4}^{B}$ and the terms with $\Phi_{22}^{A}$ disappear.
Thus under the gauge conditions \eqref{gauge1} and \eqref{gauge2}, the decoupled wave equation for $\Psi_{4}^{B}$
is obtained as
\begin{align}
&\bigl[(\Delta+3\gamma-\bar{\gamma}+4\mu+\bar{\mu})(D+4\epsilon-\rho) \notag\\
&\qquad {}-(\bar{\delta}+3\alpha+\bar{\beta}-\bar{\tau}+4\pi)(\delta+4\beta-\tau)-3\Psi_{2}+2\Phi_{11}\bigr]^{A}
\Psi_{4}^{B}=T_{4}, \label{eqpsi4}
\end{align}
One can also consider the wave equation for $\Psi_{0}^{B}$, which can be obtained in a similar way.
The resultant equation becomes
\begin{align}
&\bigl[(D-3\epsilon+\bar{\epsilon}-4\rho-\bar{\rho})(\Delta-4\gamma+\mu) \notag\\
&\qquad {}-(\delta-3\beta-\bar{\alpha}+\bar{\pi}-4\tau)(\bar{\delta}-4\alpha+\pi)-3\Psi_{2}+2\Phi_{11}\bigr]^{A}
\Psi_{0}^{B}=T_{0}. \label{eqpsi0}
\end{align}
The gauge conditions are \eqref{gauge2} and
\begin{gather}
\Xi_{00}\equiv(\delta+\bar{\pi}-2\bar{\alpha}-2\beta)^{B}\Phi_{00}^{A}=0. \label{gauge3}
\end{gather}
Note that for the case $\mathcal{D} \!=\! 1$ we have \eqref{0022}, then \eqref{gauge1} and \eqref{gauge3} are
obviously satisfied. The other conditions \eqref{gauge2} are also relaxed as just $\lambda^{B}=0$ for \eqref{eqpsi4}
and just $\sigma^{B}=0$ for \eqref{eqpsi0}.

\section{Gauge conditions}

Here we will study the consistency of the gauge conditions \eqref{gauge1}, \eqref{gauge3} and \eqref{gauge2}
(or just \eqref{gauge2} for $\mathcal{D} \!=\! 1$). In Newman-Penrose formalism, there are ten gauge degrees of freedom.
Six of them are the tetrad rotation (the local Lorentz transformation), which can be decomposed as the
following three kinds \cite{Janis:1965tx}:
\begin{align}
& l^{\mu}\to l^{\mu},\quad m^{\mu}\to m^{\mu}+al^{\mu}, \quad \bar{m}^{\mu}\to \bar{m}^{\mu}+\bar{a}l^{\mu},
\quad n^{\mu}\to n^{\mu}+\bar{a}m^{\mu}+a\bar{m}^{\mu}+a\bar{a}l^{\mu},
\label{rot01}\\
& n^{\mu}\to n^{\mu},\quad m^{\mu}\to m^{\mu}+bn^{\mu}, \quad \bar{m}^{\mu}\to \bar{m}^{\mu}+\bar{b}n^{\mu},
\quad l^{\mu}\to l^{\mu}+\bar{b}m^{\mu}+b\bar{m}^{\mu}+b\bar{b}n^{\mu},
\label{rot02}\\
& l^{\mu}\to e^{-c}l^{\mu},\quad n^{\mu}\to e^{c}n^{\mu},\quad
m^{\mu}\to e^{i\vartheta}m^{\mu},\quad \bar{m}^{\mu}\to e^{-i\vartheta}\bar{m}^{\mu}.
\label{rot03}
\end{align}
Here $a$ and $b$ are complex functions, and $c$ and $\vartheta$ are real functions.
The transformations of the quantities in Newman-Penrose equations (the spin coefficients, the Weyl scalars, etc.)
under the above are shown in \cite{Chandrasekhar:1985kt}, for example.
The other four are the general coordinate transformation $x^{\prime \mu}=x^{\mu}+\xi^{\mu}(x)$.
Under the transformation, the null tetrads behave as the one-forms or the vector fields,
and the quantities in Newman-Penrose equations behave as the scalar fields.
Because of this, the form of the transformation for the latter becomes
\begin{gather}
X\to X-\xi^{\mu}\partial_{\mu}X, \label{cotr}
\end{gather}
where $X$ represents either one of the spin coefficients, $\Psi$'s, $\Phi$'s or $\Lambda$.
Since we want to keep the background  \eqref{bg11}--\eqref{bg18}
(or \eqref{bg21}-\eqref{bg24} for $\mathcal{D} \!=\! 1$) intact, the functions $a$, $b$, $c$, $\vartheta$ and $\xi^{\mu}$
have to be at the order of the perturbed quantities, and in particular the second and the higher orders of them are neglected.

The gauge conditions  \eqref{gauge1}, \eqref{gauge2} and \eqref{gauge3} are transformed
under the the gauge transformations \eqref{rot01}, \eqref{rot02}, \eqref{rot03} and \eqref{cotr}
as\footnote{Our gauge conditions are invariant under the transformation generated by $\vartheta$. }
\begin{align}
\Xi_{00}&\to \Xi_{00}+(\delta+\bar{\pi}-2\bar{\alpha}-2\beta)^{A}(\xi^{\mu}\partial_{\mu}\Phi_{00}^{A})
+D^{A}(a\Phi_{00}^{A})-2a\rho^{A}\Phi_{00}^{A}
\notag\\
&\qquad {}+b(\Delta^{A}+\bar{\mu}^{A}-2\mu^{A}-2\bar{\gamma}^{A}-2\gamma^{A})\Phi_{00}^{A}
+2\Phi_{00}^{A}\delta c,
\\
\Xi_{22}&\to \Xi_{22}+(\bar{\delta}-\bar{\tau}+2\alpha+2\bar{\beta})^{A}(\xi^{\mu}\partial_{\mu}\Phi_{22}^{A})
+\bar{a}(D^{A}-\bar{\rho}^{A}+2\rho^{A})\Phi_{22}^{A}
\notag\\
&\qquad {}+\Delta(\bar{b}\Phi_{22}^{A})+2\bar{b}(\mu^{A}+\gamma^{A}+\bar{\gamma}^{A})\Phi_{22}^{A}
-2\Phi_{22}^{A}\bar{\delta} c,
\\
\lambda^{B}&\to \lambda^{B}+2\alpha^{A}\bar{a}+\bar{\delta}^{A}\bar{a},
\\
\sigma^{B}&\to \sigma^{B}+2\beta^{A}b-\delta^{A}b.
\end{align}
One can find that there are nine real degrees (four $\xi$'s, $a$, $b$ and $c$) of freedom for eight real conditions
($\Xi_{00}$, $\Xi_{22}$, $\lambda^{B}$ and $\sigma^{B}$). In general, it is enough to satisfy all the conditions.

In order to see the the condition explicitly, here we will use the form of the metric perturbation,
following the $A$-$K$ parametrization \cite{Thompson:2016fxe,Liu:2022csl}. First, the metric perturbation is given as
\begin{gather}
g_{\mu\nu}=g_{\mu\nu}^{A}+h_{\mu\nu}^{BE}+h_{\mu\nu}^{BO},
\end{gather}
where $g_{\mu\nu}^{A}=g_{\mu\nu}^{\mathrm{eff}}$ is the background metric \eqref{bg}.
$h_{\mu\nu}^{BE}$ and $h_{\mu\nu}^{BO}$ are respectively the even- and the odd-parity part of the perturbation,
which are parametrized as
\begin{align}
h_{\mu\nu}^{BE}&=
\begin{pmatrix}
AS & -DS & -rB\partial_{\theta}S & -rB\partial_{\varphi}S \\
   &  KS &  rH\partial_{\theta}S &  rH\partial_{\varphi}S \\
   &     &  r^{2}E_{F}^{+}       &  r^{2}F P_{S1}         \\
   &     &                       &  r^{2}\sin^{2}\theta\,E_{F}^{-}
\end{pmatrix},
\\
h_{\mu\nu}^{BO}&=\frac{1}{\mathcal{D}}
\begin{pmatrix}
0 & 0 &  rC\csc\theta\partial_{\varphi}S & -rC\sin\theta\partial_{\theta}S \\
  & 0 & -rJ\csc\theta\partial_{\varphi}S &  rJ\sin\theta\partial_{\theta}S \\
  &   & -r^{2}G\csc\theta P_{S1}         & -\frac{1}{2}r^{2}G P_{S2}       \\
  &   &                                  & r^{2}G P_{S3}
\end{pmatrix}.
\end{align}
Here the lower-left blanks have to be filled as $h_{\mu\nu}^{BE}$ and $h_{\mu\nu}^{BO}$ to be symmetric.
$A$, $B$, $C$, $D$, $E$, $F$, $G$, $H$, $J$ and $K$ are the functions\footnote{We hope the readers may not
confuse the function $D$ here and the differential operator $D$ in Newman-Penrose formalism.} of $t$ and $r$,
and $S$ is the function of $\theta$ and $\varphi$, which should in turn be identified as the spherical harmonics
$Y_{lm}(\theta, \varphi)$.
The quantities $E^{\pm}_{F}$, $P_{S1}$, $P_{S2}$ and $P_{S3}$ are defined by
\begin{align}
E^{\pm}_{F}&=\left[E\pm F\left(\partial_{\theta}^{2}+\frac{1}{2}l(l+1)\right)\right]S,
\\
P_{S1}&=(\partial_{\theta}\partial_{\varphi}-\cot\theta\partial_{\varphi})S,
\\
P_{S2}&=(\csc\theta\partial_{\varphi}^{2}+\cos\theta\partial_{\theta}-\sin\theta\partial_{\theta}^{2})S,
\\
P_{S3}&=(\sin\theta\partial_{\theta}\partial_{\varphi}-\cos\theta\partial_{\varphi})S,
\end{align}
where $l$ is one of the label in $Y_{lm}(\theta, \varphi)$.
Other perturbed quantities are also decomposed into the even- and the odd-parity modes, such as
$l_{\mu}^{B}=l_{\mu}^{BE}+l_{\mu}^{BO}$, etc.

Now we have to find the null tetrads corresponding to the metric.
Even though the metric \eqref{bg} is fixed, the tetrads are not unique due to the tetrad rotation
\eqref{rot01}, \eqref{rot02} and \eqref{rot03}. In other words, once a set of null tetrads is found,
the others are obtained from those tetrads by the tetrad rotation. We will fix the \textit{reference} tetrads as
\begin{align}
l_{\mu}^{E}dx^{\mu}&\equiv(l_{\mu}^{A}+l_{\mu}^{BE})dx^{\mu}
\notag\\
&=dt
-\frac{\mathcal{D}}{\mathcal{A}}\left[1-\frac{1}{2\mathcal{A}}
\left(A-\frac{2\mathcal{A}}{\mathcal{D}}D+\frac{\mathcal{A}^{2}}{\mathcal{D}^{2}}K\right)S\right]dr
-\frac{r}{\mathcal{A}\mathcal{D}}(\mathcal{D}B-\mathcal{A}H)dS,
\\
n_{\mu}^{E}dx^{\mu}&\equiv(n_{\mu}^{A}+n_{\mu}^{BE})dx^{\mu}
\notag\\
&=\!\frac{\mathcal{A}}{2}\!\left(\!1\!+\!\frac{AS}{\mathcal{A}}\right)dt\!+\!\frac{\mathcal{D}}{2}\left[1\!+\!\frac{1}{2\mathcal{A}}
\left(A\!-\!\frac{2\mathcal{A}}{\mathcal{D}}D\!-\!\frac{\mathcal{A}^{2}}{\mathcal{D}^{2}}K\right)S\right]dr
\!-\!\frac{r}{2\mathcal{D}}(\mathcal{D}B\!+\!\mathcal{A}H)dS,
\\
m_{\mu}^{E}dx^{\mu}&\equiv(m_{\mu}^{A}+m_{\mu}^{BE})dx^{\mu}
\notag\\
&=-\frac{r}{\sqrt{2}}\left[\left(1-\frac{E_{F}^{+}}{2}\right)d\theta
+i\left(1-\frac{E_{F}^{-}}{2}+i\csc\theta F P_{S1}\right)\sin\theta d\varphi\right],
\\
\bar{m}_{\mu}^{E}dx^{\mu}&\equiv(\bar{m}_{\mu}^{A}+\bar{m}_{\mu}^{BE})dx^{\mu}
\notag\\
&=-\frac{r}{\sqrt{2}}\left[\left(1-\frac{E_{F}^{+}}{2}\right)d\theta
-i\left(1-\frac{E_{F}^{-}}{2}-i\csc\theta F P_{S1}\right)\sin\theta d\varphi\right],
\end{align}
for the even-parity modes and
\begin{align}
l_{\mu}^{O}dx^{\mu}&\equiv(l_{\mu}^{A}+l_{\mu}^{BO})dx^{\mu}
\notag\\
&=dt-\frac{\mathcal{D}}{\mathcal{A}}dr
+\frac{r}{\mathcal{A}\mathcal{D}}\left(C-\frac{\mathcal{A}}{\mathcal{D}}J\right)
(\csc\theta\partial_{\varphi}S\,d\theta-\sin\theta\partial_{\theta}S\,d\varphi),
\\
n_{\mu}^{O}dx^{\mu}&\equiv(n_{\mu}^{A}+n_{\mu}^{BO})dx^{\mu}
\notag\\
&=\frac{\mathcal{A}}{2}dt+\frac{\mathcal{D}}{2}dr
+\frac{r}{2\mathcal{D}}\left(C+\frac{\mathcal{A}}{\mathcal{D}}J\right)
(\csc\theta\partial_{\varphi}S\,d\theta-\sin\theta\partial_{\theta}S\,d\varphi),
\\
m_{\mu}^{O}dx^{\mu}&\equiv(m_{\mu}^{A}+m_{\mu}^{BO})dx^{\mu}
\notag\\
&=-\frac{r}{\sqrt{2}}\left[\left(1\!+\!\frac{\csc\theta}{2\mathcal{D}}G P_{S1}\right)d\theta
\!+\!i\sin\theta\left(1-\frac{1}{2\mathcal{D}}\csc\theta G (P_{S1}+iP_{S2})\right)d\varphi\right],
\\
\bar{m}_{\mu}^{O}dx^{\mu}&\equiv(\bar{m}_{\mu}^{A}+\bar{m}_{\mu}^{BO})dx^{\mu}
\notag\\
&=-\frac{r}{\sqrt{2}}\left[\left(1\!+\!\frac{\csc\theta}{2\mathcal{D}}G P_{S1}\right)d\theta
\!-\!i\sin\theta\left(1-\frac{1}{2\mathcal{D}}\csc\theta G (P_{S1}-iP_{S2})\right)d\varphi\right],
\end{align}
for the odd-parity modes. Here $dS=(\partial_{\theta}S)d\theta+(\partial_{\varphi}S)d\varphi$.
Note that the above choice of the reference tetrads is different from the one taken in \cite{Jing:2021ahx}
for both the even- and the odd-parity modes even 
under the Regge-Wheeler gauge $B=F=H=G=0$ \cite{Regge:1957td,Zerilli:1970se,Moncrief:1974am}
or the EZ gauge $B=E=F=G=0$ \cite{Thompson:2016fxe,Detweiler:2008ft}.
This is again due to the different choice of the reference and they are equivalent.
The advantage of our choice is that the expressions of $\lambda^{B}$ and $\sigma^{B}$
become relatively simple as
\begin{align}
\lambda^{BE}&=
-\frac{\mathcal{A}}{4}\left\{\left[\partial_{\theta}^{2}+\frac{1}{2}l(l+1)\right]S-i\csc\theta P_{S1}\right\}
\left(\frac{1}{\mathcal{A}}\partial_{t}F-\frac{1}{\mathcal{D}}\partial_{r}F\right),
\\
\sigma^{BE}&=
\frac{1}{2}\left\{\left[\partial_{\theta}^{2}+\frac{1}{2}l(l+1)\right]S+i\csc\theta P_{S1}\right\}
\left(\frac{1}{\mathcal{A}}\partial_{t}F+\frac{1}{\mathcal{D}}\partial_{r}F\right),
\end{align}
for the even-parity modes and
\begin{align}
\lambda^{BO}&=
\frac{\mathcal{A}}{8}\csc\theta \left(2P_{S1}-iP_{S2}\right)
\left[\frac{1}{\mathcal{A}}\partial_{t}\left(\frac{G}{\mathcal{D}}\right)
-\frac{1}{\mathcal{D}}\partial_{r}\left(\frac{G}{\mathcal{D}}\right)\right],
\\
\sigma^{BO}&=
-\frac{1}{4}\csc\theta \left(2P_{S1}+iP_{S2}\right)
\left[\frac{1}{\mathcal{A}}\partial_{t}\left(\frac{G}{\mathcal{D}}\right)
+\frac{1}{\mathcal{D}}\partial_{r}\left(\frac{G}{\mathcal{D}}\right)\right],
\end{align}
for the odd-parity modes. In particular, one can easily find that they vanish under the Regge-Wheeler or the EZ gauge%
\footnote{For the even-parity modes, the choice taken in \cite{Jing:2021ahx} can also lead to $\lambda^{BE}=\sigma^{BE}=0$
under $F=0$, but not for the odd-parity mode under $G=0$.},
which also means that the above reference tetrads under those gauges can be used as the standard form of the perturbation.

Next we will consider the gauge invariants,
are obtained in \cite{Liu:2022csl,Thompson:2016fxe} as
\begin{align}
\boldsymbol{\alpha}&=J-\frac{r}{2}\mathcal{D}\partial_{r}\left(\frac{G}{\mathcal{D}}\right),
\\
\boldsymbol{\beta}&=-C-\frac{r}{2}\partial_{t}G,
\\
\boldsymbol{\chi}&=H-\frac{\mathcal{D}^{2}}{2\mathcal{A}}E-\frac{l(l+1)\mathcal{D}^{2}}{4\mathcal{A}}F
-\frac{r}{2}\partial_{r}F,
\\
\boldsymbol{\psi}&=\frac{1}{2}K-\frac{r}{4}\left(\frac{\mathcal{D}^{2}}{\mathcal{A}}\right)'E
-\frac{\mathcal{D}^{2}}{2\mathcal{A}}E-\frac{r\mathcal{D}^{2}}{2\mathcal{A}}\partial_{r}E
\notag\\
&\qquad{}
-\frac{r}{8}l(l+1)\left(\frac{\mathcal{D}^{2}}{\mathcal{A}}\right)'F
-\frac{\mathcal{D}^{2}}{4\mathcal{A}}l(l+1)F
-\frac{r\mathcal{D}^{2}}{4\mathcal{A}}l(l+1)\partial_{r}F
\\
\boldsymbol{\delta}&=D+\frac{r\mathcal{D}^{2}}{2\mathcal{A}}\partial_{t}E
+\left(\frac{r\mathcal{A}'}{\mathcal{A}}-1\right)B-r\partial_{r}B
\notag\\
&\qquad{}
-\frac{r^{2}}{2}\partial_{t}\partial_{r}F
-\left[r-\frac{r^{2}\mathcal{A}'}{2\mathcal{A}}-\frac{r\mathcal{D}^{2}}{4\mathcal{A}}l(l+1)\right]\partial_{t}F,
\\
\boldsymbol{\epsilon}&=-\frac{1}{2}A-\frac{r\mathcal{A}'}{4}E-r\partial_{t}B
-\frac{r\mathcal{A}'}{8}l(l+1)F-\frac{r^{2}}{2}\partial_{t}^{2}F.
\end{align}
The perturbed Weyl scalars $\Psi_{4}^{B}$ and $\Psi_{0}^{B}$ can be written as the
gauge-invariant combinations of the perturbation as
\begin{align}
\Psi_{4}^{BE}&=
\frac{\csc\theta}{8r^{2}\mathcal{D}^{3}}(P_{S2}+2iP_{S1})
\left[-\mathcal{A}^{2}\mathcal{D}\boldsymbol{\psi}-\mathcal{A}\mathcal{D}^{2}\boldsymbol{\delta}
+\mathcal{D}^{3}\boldsymbol{\epsilon}\right.
\notag\\
&\qquad\qquad\qquad\qquad\qquad {}\left.
-r\mathcal{A}\mathcal{D}^{2}\partial_{t}\boldsymbol{\chi}+r\mathcal{A}^{2}\mathcal{D}\partial_{r}\boldsymbol{\chi}
+\mathcal{A}^{2}(\mathcal{D}-r\mathcal{D}')\boldsymbol{\chi}
\right],
\\
\Psi_{4}^{BO}&=
-\frac{i\csc\theta}{8r\mathcal{D}}(P_{S2}+2iP_{S1})
\Biggl[-\frac{\mathcal{A}}{\mathcal{D}}\partial_{t}\boldsymbol{\alpha}
+\frac{\mathcal{A}^{2}}{\mathcal{D}^{2}}\partial_{r}\boldsymbol{\alpha}
+\frac{\mathcal{A}^{2}(\mathcal{D}-2r\mathcal{D}')}{r\mathcal{D}^{3}}\boldsymbol{\alpha}
\notag\\
&\qquad\qquad\qquad\qquad\qquad {}
+\partial_{t}\boldsymbol{\beta}-\frac{\mathcal{A}}{\mathcal{D}}\partial_{r}\boldsymbol{\beta}
+\left(\frac{\mathcal{A}\mathcal{D}'}{\mathcal{D}^{2}}-\frac{\mathcal{A}-r\mathcal{A}'}{r\mathcal{D}}\right)
\boldsymbol{\beta}\Biggr],
\\
\Psi_{0}^{BE}&=
\frac{\csc\theta}{2r^{2}\mathcal{A}^{2}\mathcal{D}^{3}}(P_{S2}-2iP_{S1})
\left[-\mathcal{A}^{2}\mathcal{D}\boldsymbol{\psi}+\mathcal{A}\mathcal{D}^{2}\boldsymbol{\delta}
+\mathcal{D}^{3}\boldsymbol{\epsilon}\right.
\notag\\
&\qquad\qquad\qquad\qquad\qquad {}\left.
+r\mathcal{A}\mathcal{D}^{2}\partial_{t}\boldsymbol{\chi}+r\mathcal{A}^{2}\mathcal{D}\partial_{r}\boldsymbol{\chi}
+\mathcal{A}^{2}(\mathcal{D}-r\mathcal{D}')\boldsymbol{\chi}
\right],
\\
\Psi_{0}^{BO}&=
\frac{i\csc\theta}{2r\mathcal{A}^{2}\mathcal{D}}(P_{S2}-2iP_{S1})
\Biggl[\frac{\mathcal{A}}{\mathcal{D}}\partial_{t}\boldsymbol{\alpha}
+\frac{\mathcal{A}^{2}}{\mathcal{D}^{2}}\partial_{r}\boldsymbol{\alpha}
+\frac{\mathcal{A}^{2}(\mathcal{D}-2r\mathcal{D}')}{r\mathcal{D}^{3}}\boldsymbol{\alpha}
\notag\\
&\qquad\qquad\qquad\qquad\qquad {}
+\partial_{t}\boldsymbol{\beta}+\frac{\mathcal{A}}{\mathcal{D}}\partial_{r}\boldsymbol{\beta}
+\left(-\frac{\mathcal{A}\mathcal{D}'}{\mathcal{D}^{2}}+\frac{\mathcal{A}-r\mathcal{A}'}{r\mathcal{D}}\right)
\boldsymbol{\beta}\Biggr],
\end{align}
The gauge conditions \eqref{gauge1}, \eqref{gauge3} and \eqref{gauge2} can be expressed as the gauge invariants
and the set of the variables $(B,F,H,G)$ or $(B,E,F,G)$. The explicit form for the latter is shown in Appendix.

\section{Explicit PDE and ODE for radial coordinate}

In section 3, we have obtained the wave equation for the perturbed Weyl scalars $\Psi_{4}^{B}$ and $\Psi_{0}^{B}$
as \eqref{eqpsi4} and \eqref{eqpsi0}, respectively. By substituting the background \eqref{bg11}--\eqref{bg18}
(or \eqref{bg21}-\eqref{bg24} for $\mathcal{D} \!=\! 1$), the partial differential equations (the master equations) are
obtained, Then by the separation of variables, the ordinary differential equation for the radial coordinate $r$
and that for the angular coordinates $\theta$ and $\varphi$ are also obtained.

We define $\psi_{(-2)}$ as
\begin{gather}
\psi_{(-2)}\equiv(\rho^{A})^{-4}\Psi_{4}^{B}=(r\mathcal{D})^{4}\Psi_{4}^{B}.
\end{gather}
From \eqref{eqpsi4}, $\psi_{(-2)}$  satisfies the following partial differential equation:
\begin{align}
&\frac{r^{2}}{\mathcal{A}}\frac{\partial^{2}\psi_{(-2)}}{\partial t^{2}}
-(r^{2}\mathcal{A})^{2}\mathcal{D}^{7}\frac{\partial}{\partial r}
\left[(r^{2}\mathcal{A})^{-1}\mathcal{D}^{-9}\frac{\partial\psi_{(-2)}}{\partial r}\right]
+\left(\frac{2r^{2}\mathcal{A}'}{\mathcal{A}\mathcal{D}}-\frac{4r}{\mathcal{D}}\right)
\frac{\partial\psi_{(-2)}}{\partial t}
\notag\\
&\qquad {}+\left[-\frac{24r^{2}\mathcal{A}(\mathcal{D}')^{2}}{\mathcal{D}^{4}}
+\frac{4r^{2}\mathcal{A}\mathcal{D}''}{\mathcal{D}^{3}}-\frac{3r^{2}\mathcal{A}'\mathcal{D}'}{\mathcal{D}^{3}}
-\frac{12r\mathcal{A}\mathcal{D}'}{\mathcal{D}^{3}}-\frac{r^{2}\mathcal{A}''}{\mathcal{D}^{2}}
-\frac{2r\mathcal{A}'}{\mathcal{D}^{2}}\right]\psi_{(-2)}
\notag\\
&\qquad {}\!-\!\frac{1}{\sin\theta}\frac{\partial}{\partial\theta}\left(\sin\theta\frac{\partial\psi_{(-2)}}{\partial\theta}\right)
\!-\!\frac{1}{\sin^{2}\theta}\frac{\partial^{2}\psi_{(-2)}}{\partial\varphi^{2}}
\!+\!\frac{4i\cot\theta}{\sin\theta}\frac{\partial\psi_{(-2)}}{\partial\varphi}\!+\!(4\cot^{2}\theta\!+\!2)\psi_{(-2)}
\notag\\
&\qquad {}=2r^{6}\mathcal{D}^{4}T_{4}. \label{pde4}
\end{align}
For homogeneous case ($T_{4}=0$), by assuming the product form
$\psi_{(-2)}=e^{-i\omega t}e^{im\varphi}R(r)\mathcal{S}(\theta)$,
one can obtain the separated equations as
\begin{align}
&(r^{2}\mathcal{A})^{2}\mathcal{D}^{7}\frac{d}{dr}\left[(r^{2}\mathcal{A})^{-1}\mathcal{D}^{-9}\frac{dR(r)}{dr}\right]
+\left[\frac{r^{2}\omega^{2}}{\mathcal{A}}
+i\omega\left(\frac{2r^{2}\mathcal{A}'}{\mathcal{A}\mathcal{D}}-\frac{4r}{\mathcal{D}}\right)+\frac{24r^{2}\mathcal{A}(\mathcal{D}')^{2}}{\mathcal{D}^{4}}\right.
\notag\\
&\qquad {}\left.
-\frac{4r^{2}\mathcal{A}\mathcal{D}''}{\mathcal{D}^{3}}
+\frac{3r^{2}\mathcal{A}'\mathcal{D}'}{\mathcal{D}^{3}}
+\frac{12r\mathcal{A}\mathcal{D}'}{\mathcal{D}^{3}}
+\frac{r^{2}\mathcal{A}''}{\mathcal{D}^{2}}
+\frac{2r\mathcal{A}'}{\mathcal{D}^{2}}
-\boldsymbol{\lambda}\right]R(r)=0,
\label{req01}
\\
&\frac{1}{\sin\theta}\frac{d}{d\theta}\left(\sin\theta\frac{d \mathcal{S}(\theta)}{d\theta}\right)
-\left(\frac{m^{2}}{\sin^{2}\theta}-\frac{4m\cot\theta}{\sin\theta}+4\cot^{2}\theta+2
-\boldsymbol{\lambda}\right)\mathcal{S}(\theta)=0,
\label{angeq01}
\end{align}
where $\omega$ gives the frequency of the gravitational wave and $\boldsymbol{\lambda}$ is the separation constant.
From \eqref{angeq01} one can find that $S(\theta)e^{im\varphi}$ coincides with
the $s=-2$ spin-weighted spherical harmonics, denoted as ${}_{-2}Y_{lm}(\theta,\varphi)$,
and then the separation constant $\boldsymbol{\lambda}$ has to be
\begin{align}
\boldsymbol{\lambda}=(l-1)(l+2),
\end{align}
where $l$ and $m$ take the integer value with $l \ge 2$ and $-l \le m \le l$, respectively.
For inhomogeneous case ($T_{4}\neq 0$), one can expand $\psi_{(-2)}$ and $T_{4}$ as
\begin{align}
\psi_{(-2)}&=\int d\omega \sum_{l,m}R_{(-2)l\omega}(r){}_{-2}Y_{lm}(\theta,\varphi)e^{-i\omega t}, \\
2r^{6}\mathcal{D}^{4}T_{4}&=-\int d\omega \sum_{l,m}G_{(-2)l\omega}(r){}_{-2}Y_{lm}(\theta,\varphi)e^{-i\omega t}.
\end{align}
Then the ordinary differential equation for the radial coordinate $r$ is
\begin{align}
&(r^{2}\mathcal{A})^{2}\mathcal{D}^{7}\frac{d}{dr}
\left[(r^{2}\mathcal{A})^{-1}\mathcal{D}^{-9}\frac{dR_{(-2)l\omega}(r)}{dr}\right]
+\left[\frac{r^{2}\omega^{2}}{\mathcal{A}}
+i\omega\left(\frac{2r^{2}\mathcal{A}'}{\mathcal{A}\mathcal{D}}-\frac{4r}{\mathcal{D}}\right)\right.
\notag\\
&\qquad {}\left.
+\frac{24r^{2}\mathcal{A}(\mathcal{D}')^{2}}{\mathcal{D}^{4}}
-\frac{4r^{2}\mathcal{A}\mathcal{D}''}{\mathcal{D}^{3}}
+\frac{3r^{2}\mathcal{A}'\mathcal{D}'}{\mathcal{D}^{3}}
+\frac{12r\mathcal{A}\mathcal{D}'}{\mathcal{D}^{3}}
+\frac{r^{2}\mathcal{A}''}{\mathcal{D}^{2}}
+\frac{2r\mathcal{A}'}{\mathcal{D}^{2}}\right.
\notag\\
&\qquad {}
-(l-1)(l+2)\biggr]R_{(-2)l\omega}(r)=G_{(-2)l\omega}(r).
\label{req02}
\end{align}
In the case of $\mathcal{D} \!=\! 1$, \eqref{req02} is reduced to
\begin{align}
&(r^{2}\mathcal{A})^{2}\frac{d}{dr}\left[(r^{2}\mathcal{A})^{-1}\frac{dR_{(-2)l\omega}(r)}{dr}\right]
\notag\\
&
\qquad{}+\left[\frac{r^{2}\omega^{2}}{\mathcal{A}}+i\omega\left(\frac{2r^{2}\mathcal{A}'}{\mathcal{A}}-4r\right)
+r^{2}\mathcal{A}''+2r\mathcal{A}'-(l-1)(l+2)\right]R_{(-2)l\omega}(r)
\notag\\
&\qquad {}
=G_{(-2)l\omega}(r),
\label{req03}
\end{align}
Note that for the case of Schwarzschild background, \textit{i.\,e.} $\mathcal{A}=1-2M/r$,
the differential equation \eqref{req03}
is reduced to the Teukolsky equation \cite{Teukolsky:1973ha} with the spin weight $s=-2$.

The equation for $\psi_{(2)}\equiv\Psi_{0}^{B}$ can be obtained in a similar way.
The master equation becomes
\begin{align}
&\frac{r^{2}}{\mathcal{A}}\frac{\partial^{2}\psi_{(2)}}{\partial t^{2}}
-(r^{2}\mathcal{A})^{-2}\mathcal{D}^{-1}\frac{\partial}{\partial r}
\left[(r^{2}\mathcal{A})^{3}\mathcal{D}^{-1}\frac{\partial\psi_{(2)}}{\partial r}\right]
-\left(\frac{2r^{2}\mathcal{A}'}{\mathcal{A}\mathcal{D}}-\frac{4r}{\mathcal{D}}\right)
\frac{\partial\psi_{(2)}}{\partial t}
\notag\\
&\qquad {}+\left[\frac{3r^{2}\mathcal{A}'\mathcal{D}'}{\mathcal{D}^{3}}
-\frac{3r^{2}\mathcal{A}''}{\mathcal{D}^{2}}
-\frac{10r\mathcal{A}'}{\mathcal{D}^{2}}
-\frac{4\mathcal{A}}{\mathcal{D}^{2}}\right]\psi_{(2)}
\notag\\
&\qquad {}\!-\!\frac{1}{\sin\theta}\frac{\partial}{\partial\theta}\left(\sin\theta\frac{\partial\psi_{(2)}}{\partial\theta}\right)
\!-\!\frac{1}{\sin^{2}\theta}\frac{\partial^{2}\psi_{(2)}}{\partial\varphi^{2}}
\!-\!\frac{4i\cot\theta}{\sin\theta}\frac{\partial\psi_{(2)}}{\partial\varphi}\!+\!(4\cot^{2}\theta+2)\psi_{(2)}
\notag\\
&\qquad {}=2r^{2}T_{0}. \label{pde0}
\end{align}
After the separation of the variables, for the angular part,
we have the $s=2$ spin-weighted spherical harmonics ${}_{2}Y_{lm}(\theta,\varphi)$. For the radial part,
we have the following equation:
\begin{align}
&(r^{2}\mathcal{A})^{-2}\mathcal{D}^{-1}\frac{d}{d r}
\left[(r^{2}\mathcal{A})^{3}\mathcal{D}^{-1}\frac{dR_{(2)l\omega}(r)}{d r}\right]
\notag\\
&\qquad {}+\left[\frac{r^{2}\omega^{2}}{\mathcal{A}}
-i\omega\left(\frac{2r^{2}\mathcal{A}'}{\mathcal{A}\mathcal{D}}-\frac{4r}{\mathcal{D}}\right)
-\frac{3r^{2}\mathcal{A}'\mathcal{D}'}{\mathcal{D}^{3}}
+\frac{3r^{2}\mathcal{A}''}{\mathcal{D}^{2}}
+\frac{10r\mathcal{A}'}{\mathcal{D}^{2}}
+\frac{4\mathcal{A}}{\mathcal{D}^{2}}\right.
\notag\\
&\qquad {}
-(l-2)(l+3)\biggr]R_{(2)l\omega}(r)=G_{(2)l\omega}(r),
\label{req04}
\end{align}
which is reduced to the Teukolsky equation with $s=\pm 2$
for $\mathcal{A}=1-2M/r$ and $D \equiv 1$.

\section{Summary and discussion}

In this paper, we have studied the wave equations for the perturbed Weyl scalars $\Psi_{4}^{B}$ and $\Psi_{0}^{B}$
in the background of the EOB dynamics for the spinless binary. We also obtained the Teukolsky-like equations for the
function of the radial coordinate $r$.
In the previous work \cite{Jing:2021ahx}, the special case $\mathcal{D} \!=\! 1$ is studied and the (decoupled)
wave equation is obtained only for the even-parity mode. On the other hand, here we have studied the odd-parity mode also,
and have obtained the same form of the wave equation. Moreover, we have considered a more general case including $\mathcal{D} \!\neq\! 1$.
The different form of the wave equation is proposed in \cite{Jing:2022vks} by the use of the different gauge,
although it is again restricted to the case of $\mathcal{D} \!=\! 1$. It would be interesting to find the relation
between our equations and their ones.

One can easily find that the background is assumed to be just spherically symmetric and general enough. Even in the case
$\mathcal{D} \!=\! 1$, it contains many kinds of the black holes.
A similar wave equations in the spherically symmetric background are obtained using the metric perturbation in
\cite{Liu:2022csl}, which are resemble to the Regge-Wheeler and the Zerilli equations \cite{Regge:1957td,Zerilli:1970se},
and are different from the ones obtained here.
This is because we have used the different gauge and the different master variables. However in the case of
Schwarzschild background, there is a transformation originated from the connection of the different gauges, which is called
the Chandrasekhar transformation \cite{Chandrasekhar:1975}. Moreover the relation between the $R_{(-2)l\omega}(r)$
in \eqref{req02} and $R_{(2)l\omega}(r)$ in \eqref{req04} can also be regarded as the special case
of the Chandrasekhar transformation \cite{Teukolsky:1973ha,Starobinsky1}.
It would be interesting to find a similar transformation in the present case.

Another possible generalization would be to include the effect of the spin,
where the background becomes axially symmetric and contains the Kerr black hole as an example.
In this case the method of the metric perturbation is difficult to perform, because it relies on the spherical symmetry
and the expansion by the spherical harmonics. However
the Newman-Penrose formalism would still be powerful enough, and it can be expected that
the gravitational-wave equation could be obtained.
Studying the equations for the electromagnetic and the scalar waves is also interesting.

\section*{Acknowledgements}
This work was supported in part by the National Natural Science Foundation of China (Grant No. 11973025).

\begin{appendix}

\section{Gauge conditions with gauge invariants}

The gauge conditions \eqref{gauge1}, \eqref{gauge3} and \eqref{gauge2} can be written
in terms of the gauge invariants and the variables $(B,E,F,G)$.
Each condition consists of the even-parity part $(E)$, odd-parity part $(O)$ and the transformation part $(T)$.
\begin{align}
\Xi_{22}&=\Xi_{22}^{E}+\Xi_{22}^{O}+\Xi_{22}^{T},
\\
\Xi_{22}^{E}&=\frac{\mathcal{A}^{2}(\partial_{\theta}S-i\csc\theta\partial_{\varphi}S)}{4\sqrt{2}r}
\left\{-\frac{\mathcal{A}\mathcal{D}''}{\mathcal{D}^{5}}
\left[\boldsymbol{\chi}+\frac{r}{2}\partial_{r}F+\frac{l(l+1)}{4}\frac{\mathcal{D}^{2}}{\mathcal{A}}F
+\frac{\mathcal{D}^{2}}{2\mathcal{A}}E\right]\right.
\notag\\
&\qquad{}
\!-\!\frac{\mathcal{D}'}{\mathcal{D}^{3}}\left[-\frac{1}{\mathcal{D}}\partial_{t}\chi
\!+\!\frac{\mathcal{A}}{2\mathcal{D}^{2}}\partial_{r}\chi
\!+\!\left(\!-\frac{\mathcal{A}}{2r\mathcal{D}^{2}}\!+\!\frac{3\mathcal{A}'}{2\mathcal{D}^{2}}
\!-\!\frac{7\mathcal{A}\mathcal{D}'}{2\mathcal{D}^{3}}\right)\chi\!+\!\frac{3}{2r\mathcal{A}}\boldsymbol{\epsilon}
\!-\!\frac{\mathcal{A}}{2r\mathcal{D}^{2}}\boldsymbol{\psi}
\right.
\notag\\
&\qquad{}
+\frac{1}{r\mathcal{D}}\boldsymbol{\delta}
+\frac{2}{\mathcal{A}}\partial_{t}B-\frac{\mathcal{D}}{\mathcal{A}}\partial_{t}E
+\left(\frac{\mathcal{A}'}{\mathcal{A}}-\frac{3\mathcal{D}'}{2\mathcal{D}}-\frac{1}{2r}\right)E
+\frac{3r}{4\mathcal{A}}\partial_{t}^{2}F+\frac{r\mathcal{A}}{4\mathcal{D}^{2}}\partial_{r}^{2}F
\notag\\
&\qquad{}
+\left(\frac{1}{\mathcal{D}}-\frac{l(l+1)}{2}\frac{\mathcal{D}}{\mathcal{A}}
-\frac{r\mathcal{A}'}{2\mathcal{A}\mathcal{D}}\right)\partial_{t}F
+\left(\frac{3r\mathcal{A}'}{4\mathcal{D}^{2}}-\frac{7r\mathcal{A}\mathcal{D}'}{4\mathcal{D}^{3}}\right)\partial_{r}F
\notag\\
&\qquad{}\left.\left.
-\left(\frac{1}{r}+\frac{3\mathcal{D}'}{\mathcal{D}}-\frac{2\mathcal{A}'}{\mathcal{A}}\right)\frac{l(l+1)}{4}F\right]\right\},
\\
\Xi_{22}^{O}&=\frac{-i\mathcal{A}^2(\partial_{\theta}S-i\csc\theta\partial_{\varphi}S)}{4\sqrt{2}r}
\left\{\!-\!\frac{\mathcal{A}\mathcal{D}''}{\mathcal{D}^{6}}
\left(\boldsymbol{\alpha}+\frac{r\mathcal{D}}{2}\partial_{r}\tilde{G}\right)
+\frac{\mathcal{D}'}{\mathcal{D}^{3}}\left[\frac{1}{\mathcal{D}^{2}}\partial_{t}\boldsymbol{\alpha}
-\frac{\mathcal{A}}{2\mathcal{D}^{3}}\partial_{r}\boldsymbol{\alpha}
\right.\right.
\notag\\
&\qquad{}
\!+\!\left(\frac{\mathcal{A}}{2r\mathcal{D}^{3}}\!-\!\frac{3\mathcal{A}'}{2\mathcal{D}^{3}}
\!+\!\frac{4\mathcal{A}\mathcal{D}'}{\mathcal{D}^{4}}\right)\boldsymbol{\alpha}
\!+\!\frac{1}{2\mathcal{A}\mathcal{D}}\partial_{t}\boldsymbol{\beta}\!-\!\frac{1}{\mathcal{D}^{2}}\partial_{r}\boldsymbol{\beta}
\!+\!\left(\frac{\mathcal{A}'}{\mathcal{A}\mathcal{D}^{2}}\!-\!\frac{1}{r\mathcal{D}^{2}}\!+\!\frac{\mathcal{D}'}{\mathcal{D}^{3}}\right)
\boldsymbol{\beta}
\notag\\
&\qquad{}\left.\left.
\!+\frac{r}{4\mathcal{A}}\partial_{t}^{2}\tilde{G}
+\!\left(\frac{r\mathcal{A}'}{2\mathcal{A}\mathcal{D}}-\frac{1}{\mathcal{D}}\right)\partial_{t}\tilde{G}
-\frac{r\mathcal{A}}{4\mathcal{D}^{2}}\partial_{r}^{2}\tilde{G}
+\!\left(\!-\frac{3r\mathcal{A}'}{4\mathcal{D}^{2}}+\frac{7r\mathcal{A}\mathcal{D}'}{4\mathcal{D}^{3}}\right)
\partial_{r}\tilde{G}\right]\right\},
\\
\Xi_{22}^{T}&=
\frac{\mathcal{A}}{4r\mathcal{D}^{4}}\left[\frac{\mathcal{A}\mathcal{D}'}{r}-\mathcal{A}'\mathcal{D}'
+\frac{3\mathcal{A}(\mathcal{D}')^{2}}{\mathcal{D}}-\mathcal{A}\mathcal{D}''\right]
\left(\bar{a}-\frac{\mathcal{A}}{2}\bar{b}\right)
\notag\\
&\qquad{}
\!-\!\frac{\mathcal{A}^2\mathcal{D}'}{4r\mathcal{D}^{3}}\left(\frac{\mathcal{A}'}{\mathcal{A}\mathcal{D}}\bar{a}
\!-\!\frac{1}{r\mathcal{D}}\bar{a}\!-\!\frac{\mathcal{A}}{r\mathcal{D}}\bar{b}
\!+\!\frac{1}{2}\partial_{t}\bar{b}-\frac{\mathcal{A}}{2\mathcal{D}}\partial_{r}\bar{b}
-\frac{\sqrt{2}}{r}\partial_{\theta}c+\frac{\sqrt{2}i\csc\theta}{r}\partial_{\varphi}c\right),
\\
\Xi_{00}&=\Xi_{00}^{E}+\Xi_{00}^{O}+\Xi_{00}^{T},
\\
\Xi_{00}^{E}&=\frac{\partial_{\theta}S+i\csc\theta\partial_{\varphi}S}{\sqrt{2}r}
\left\{-\frac{\mathcal{A}\mathcal{D}''}{\mathcal{D}^{5}}
\left[\boldsymbol{\chi}+\frac{r}{2}\partial_{r}F+\frac{l(l+1)}{4}\frac{\mathcal{D}^{2}}{\mathcal{A}}F
+\frac{\mathcal{D}^{2}}{2\mathcal{A}}E\right]\right.
\notag\\
&\qquad{}
\!-\!\frac{\mathcal{D}'}{\mathcal{D}^{3}}\left[\frac{1}{\mathcal{D}}\partial_{t}\chi
+\frac{\mathcal{A}}{2\mathcal{D}^{2}}\partial_{r}\chi
\!+\!\left(\frac{3\mathcal{A}'}{2\mathcal{D}^{2}}-\frac{\mathcal{A}}{2r\mathcal{D}^{2}}
-\frac{7\mathcal{A}\mathcal{D}'}{2\mathcal{D}^{3}}\right)\chi\!-\!\frac{5}{2r\mathcal{A}}\boldsymbol{\epsilon}
\!-\!\frac{\mathcal{A}}{2r\mathcal{D}^{2}}\boldsymbol{\psi}
\right.
\notag\\
&\qquad{}
-\frac{1}{r\mathcal{D}}\boldsymbol{\delta}
-\frac{2}{\mathcal{A}}\partial_{t}B+\frac{\mathcal{D}}{\mathcal{A}}\partial_{t}E
-\left(\frac{3\mathcal{D}'}{2\mathcal{D}}+\frac{1}{2r}\right)E
-\frac{5r}{4\mathcal{A}}\partial_{t}^{2}F+\frac{r\mathcal{A}}{4\mathcal{D}^{2}}\partial_{r}^{2}F
\notag\\
&\qquad{}
+\left(-\frac{1}{\mathcal{D}}+\frac{l(l+1)}{2}\frac{\mathcal{D}}{\mathcal{A}}
+\frac{r\mathcal{A}'}{2\mathcal{A}\mathcal{D}}\right)\partial_{t}F
+\left(\frac{3r\mathcal{A}'}{4\mathcal{D}^{2}}-\frac{7r\mathcal{A}\mathcal{D}'}{4\mathcal{D}^{3}}\right)\partial_{r}F
\notag\\
&\qquad{}\left.\left.
-\left(\frac{1}{r}+\frac{3\mathcal{D}'}{\mathcal{D}}\right)\frac{l(l+1)}{4}F\right]\right\},
\\
\Xi_{00}^{O}&=\frac{i(\partial_{\theta}S+i\csc\theta\partial_{\varphi}S)}{\sqrt{2}r}
\left\{-\frac{\mathcal{A}\mathcal{D}''}{\mathcal{D}^{6}}
\left(\boldsymbol{\alpha}+\frac{r\mathcal{D}}{2}\partial_{r}\tilde{G}\right)
-\frac{\mathcal{D}'}{\mathcal{D}^{3}}\left[\frac{1}{\mathcal{D}^{2}}\partial_{t}\boldsymbol{\alpha}
+\frac{\mathcal{A}}{2\mathcal{D}^{3}}\partial_{r}\boldsymbol{\alpha}
\right.\right.
\notag\\
&\qquad{}
\!+\!\left(\frac{3\mathcal{A}'}{2\mathcal{D}^{3}}\!-\!\frac{\mathcal{A}}{2r\mathcal{D}^{3}}
\!-\!\frac{4\mathcal{A}\mathcal{D}'}{\mathcal{D}^{4}}\right)\boldsymbol{\alpha}
\!-\!\frac{1}{2\mathcal{A}\mathcal{D}}\partial_{t}\boldsymbol{\beta}\!-\!\frac{1}{\mathcal{D}^{2}}\partial_{r}\boldsymbol{\beta}
\!+\!\left(\frac{\mathcal{A}'}{\mathcal{A}\mathcal{D}^{2}}\!-\!\frac{1}{r\mathcal{D}^{2}}\!+\!\frac{\mathcal{D}'}{\mathcal{D}^{3}}\right)
\boldsymbol{\beta}
\notag\\
&\qquad{}\left.\left.
-\frac{r}{4\mathcal{A}}\partial_{t}^{2}\tilde{G}
+\left(\frac{r\mathcal{A}'}{2\mathcal{A}\mathcal{D}}-\frac{1}{\mathcal{D}}\right)\partial_{t}\tilde{G}
+\frac{r\mathcal{A}}{4\mathcal{D}^{2}}\partial_{r}^{2}\tilde{G}
+\left(\frac{3r\mathcal{A}'}{4\mathcal{D}^{2}}-\frac{7r\mathcal{A}\mathcal{D}'}{4\mathcal{D}^{3}}\right)
\partial_{r}\tilde{G}\right]\right\},
\\
\Xi_{00}^{T}&=
\frac{1}{r\mathcal{D}^{4}}\left[\frac{\mathcal{D}'}{r}
+\frac{3(\mathcal{D}')^{2}}{\mathcal{D}}-\mathcal{D}''\right]
\left(a-\frac{\mathcal{A}}{2}b\right)
\notag\\
&\qquad{}
\!-\!\frac{\mathcal{D}'}{r\mathcal{D}^{3}}\left(\frac{2}{r\mathcal{D}}a
\!+\!\frac{1}{\mathcal{A}}\partial_{t}a+\frac{1}{\mathcal{D}}\partial_{r}a
+\frac{\mathcal{A}}{2r\mathcal{D}}b-\frac{\mathcal{A}'}{\mathcal{D}}b
+\frac{\sqrt{2}}{r}\partial_{\theta}c+\frac{\sqrt{2}i\csc\theta}{r}\partial_{\varphi}c\right),
\\
\lambda^{B}&=\lambda^{BE}+\lambda^{BO}+\lambda^{BT},
\\
\lambda^{BE}&=
-\frac{\mathcal{A}}{4}\left\{\left[\partial_{\theta}^{2}+\frac{1}{2}l(l+1)\right]S-i\csc\theta P_{S1}\right\}
\left(\frac{1}{\mathcal{A}}\partial_{t}F-\frac{1}{\mathcal{D}}\partial_{r}F\right),
\\
\lambda^{BO}&=
\frac{\mathcal{A}}{8}\csc\theta \left(2P_{S1}-iP_{S2}\right)
\left(\frac{1}{\mathcal{A}}\partial_{t}\tilde{G}-\frac{1}{\mathcal{D}}\partial_{r}\tilde{G}\right),
\\
\lambda^{BT}&=
\frac{1}{\sqrt{2}r}(\partial_{\theta}\bar{a}-i\csc\theta\partial_{\varphi}\bar{a}-\bar{a}\cot\theta),
\\
\sigma^{B}&=\sigma^{BE}+\sigma^{BO}+\sigma^{BT},
\\
\sigma^{BE}&=
\frac{1}{2}\left\{\left[\partial_{\theta}^{2}+\frac{1}{2}l(l+1)\right]S+i\csc\theta P_{S1}\right\}
\left(\frac{1}{\mathcal{A}}\partial_{t}F+\frac{1}{\mathcal{D}}\partial_{r}F\right),
\\
\sigma^{BO}&=
-\frac{1}{4}\csc\theta \left(2P_{S1}+iP_{S2}\right)
\left(\frac{1}{\mathcal{A}}\partial_{t}\tilde{G}+\frac{1}{\mathcal{D}}\partial_{r}\tilde{G}\right),
\\
\sigma^{BT}&=
-\frac{1}{\sqrt{2}r}(\partial_{\theta}b+i\csc\theta\partial_{\varphi}b-b\cot\theta).
\end{align}
Here $\tilde{G}=G/\mathcal{D}$ and $\xi$'s are omitted since these are converted to fix the variables $(B,E,F,G)$.
Note that in the case of $\mathcal{D}\!=\! 1$, the conditions $\Xi_{22}=\Xi_{00}=0$ becomes trivial from \eqref{0022}
and $\lambda^{B}=\sigma^{B}=0$ can be achieved by setting $F=G=a=b=0$.

\end{appendix}


\begin{thebibliography}{99}

\bibitem{Abbott:2016blz}
B.~Abbott \textit{et al.} [LIGO Scientific and Virgo],
Phys. Rev. Lett. \textbf{116}, no.6, 061102 (2016)
doi:10.1103/PhysRevLett.116.061102
[arXiv:1602.03837 [gr-qc]].

\bibitem{PoissonWill}
E.~Poisson and C.~M.~Will,
``Gravity: Newtonian, Post-Newtonian, Relativistic.''
Cambridge University Press (2014)
doi:10.1017/CBO9781139507486.

\bibitem{Mino:1997bx}
Y.~Mino, M.~Sasaki, M.~Shibata, H.~Tagoshi and T.~Tanaka,
Prog. Theor. Phys. Suppl. \textbf{128}, 1-121 (1997)
doi:10.1143/PTPS.128.1
[arXiv:gr-qc/9712057 [gr-qc]].

\bibitem{Fujita:2014eta}
R.~Fujita,
PTEP \textbf{2015}, no.3, 033E01 (2015)
doi:10.1093/ptep/ptv012
[arXiv:1412.5689 [gr-qc]].

\bibitem{Buonanno:2000ef}
A.~Buonanno and T.~Damour,
Phys. Rev. D \textbf{62}, 064015 (2000)
doi:10.1103/PhysRevD.62.064015
[arXiv:gr-qc/0001013 [gr-qc]].

\bibitem{Damour:2016gwp}
T.~Damour,
Phys. Rev. D \textbf{94}, no.10, 104015 (2016)
doi:10.1103/PhysRevD.94.104015
[arXiv:1609.00354 [gr-qc]].

\bibitem{Jing:2021ahx}
J.~Jing, S.~Chen, M.~Sun, X.~He, M.~Wang and J.~Wang,
Sci. China Phys. Mech. Astron. \textbf{65}, no.6, 260411 (2022)
doi:10.1007/s11433-022-1885-6
[arXiv:2112.09838 [gr-qc]].

\bibitem{Newman:1961qr}
  E.~Newman and R.~Penrose,
  J.\ Math.\ Phys.\  {\bf 3}, 566 (1962)
  doi:10.1063/1.1724257.

\bibitem{Teukolsky:1973ha}
  S.~A.~Teukolsky,
  Astrophys.\ J.\  {\bf 185}, 635 (1973)
  doi:10.1086/152444.

\bibitem{Jing:2022vks}
J.~Jing, S.~Long, W.~Deng, M.~Wang and J.~Wang,
Sci. China Phys. Mech. Astron. \textbf{65}, no.10, 100411 (2022)
doi:10.1007/s11433-022-1951-1
[arXiv:2208.02420 [gr-qc]].

\bibitem{Lenzi:2021wpc}
M.~Lenzi and C.~F.~Sopuerta,
Phys. Rev. D \textbf{104}, no.8, 084053 (2021)
doi:10.1103/PhysRevD.104.084053
[arXiv:2108.08668 [gr-qc]].

\bibitem{Liu:2022csl}
W.~Liu, X.~Fang, J.~Jing and A.~Wang,
Sci. China Phys. Mech. Astron. \textbf{66}, no.1, 210411 (2023)
doi:10.1007/s11433-022-1956-4
[arXiv:2201.01259 [gr-qc]].

\bibitem{Janis:1965tx}
A.~I.~Janis and E.~T.~Newman,
J. Math. Phys. \textbf{6}, 902-914 (1965)
doi:10.1063/1.1704349

\bibitem{Chandrasekhar:1985kt}
S.~Chandrasekhar,
``The mathematical theory of black holes,''
Springer (1985)
doi:10.1007/978-94-009-6469-3\_2

\bibitem{Thompson:2016fxe}
J.~E.~Thompson, B.~F.~Whiting and H.~Chen,
Class. Quant. Grav. \textbf{34}, no.17, 174001 (2017)
doi:10.1088/1361-6382/aa7f5b
[arXiv:1611.06214 [gr-qc]].

\bibitem{Regge:1957td}
  T.~Regge and J.~A.~Wheeler,
  Phys.\ Rev.\  {\bf 108}, 1063 (1957)
  doi:10.1103/PhysRev.108.1063.

\bibitem{Zerilli:1970se}
  F.~J.~Zerilli,
  Phys.\ Rev.\ Lett.\  {\bf 24}, 737 (1970)
  doi:10.1103/PhysRevLett.24.73.7.

\bibitem{Moncrief:1974am}
  V.~Moncrief,
  Annals Phys.\  {\bf 88}, 323 (1974)
  doi:10.1016/0003-4916(74)90173-0.

\bibitem{Detweiler:2008ft}
S.~L.~Detweiler,
Phys. Rev. D \textbf{77}, 124026 (2008)
doi:10.1103/PhysRevD.77.124026
[arXiv:0804.3529 [gr-qc]].











\bibitem{Chandrasekhar:1975}
  S.~Chandrasekhar,
   Proc. R. Soc. Lond. A \textbf{A343}, 289-298 (1975)
   doi:10.1098/rspa.1975.0066

\bibitem{Starobinsky1}
A. A. Starobinsky and S. M. Churilov, Zh. Eksp. Teor. Fiz. \textbf{65}, 3,
A. A. Starobinsky and S. M. Churilov, Sov. Phys. JETP 38, 1 (1974).


%
%
%
%
%
%
%
%
%
%
%
%
%
%
%
%
%
%
%
%
%
%
%
%
%
%
%
%
%
%
%
%
%
%
%
%
%
%
%
%
%



\end{thebibliography}
\end{document}